# An Optimized Weighted Association Rule Mining On Dynamic Content

P.Velvadivu[1] and Dr.K.Duraisamy[2]

[1]Lecturer,
Department of Computer Technology and Applications,
Coimbatore Institute of Technology, Coimbatore.

[2]Dean,
School of Science and Humanities,
Kongu Engineering College,
Perundurai, Erode

**Abstract**
Association rule mining aims to explore large transaction databases for association rules. Classical Association Rule Mining (ARM) model assumes that all items have the same significance without taking their weight into account. It also ignores the difference between the transactions and importance of each and every itemsets. But, the Weighted Association Rule Mining (WARM) does not work on databases with only binary attributes. It makes use of the importance of each itemset and transaction. WARM requires each item to be given weight to reflect their importance to the user. The weights may correspond to special promotions on some products, or the profitability of different items.
This research work first focused on a weight assignment based on a directed graph where nodes denote items and links represent association rules. A generalized version of HITS is applied to the graph to rank the items, where all nodes and links are allowed to have weights. This research then uses enhanced HITS algorithm by developing an online eigenvector calculation method that can compute the results of mutual reinforcement voting in case of frequent updates. For Example in Share Market Shares price may go down or up. So we need to carefully watch the market and our association rule mining has to produce the items that have undergone frequent changes. These are done by estimating the upper bound of perturbation and postponing of the updates whenever possible. Next we prove that enhanced algorithm is more efficient than the original HITS under the context of dynamic data.

## 1. Introduction

Association Rule Mining aims to explore large transaction databases for association rules, which may reveal the implicit relationships among the data attributes [1]. It has number of practical applications, including classification, text mining, Web log analysis, Share Market and recommendation systems. The classical model of association rule mining employs the support measure, which treats every transaction equally.
In contrast, different transactions have different weights in real-life data sets. For example, in the market basket data, each transaction is recorded with some profit. Much effort has been dedicated to association rule mining with preassigned weights. However, most data types do not come with such preassigned weights, such as Web site click-stream data. There should be some notion of importance in those data. For instance, transactions with a large amount of items should be considered more important than transactions with only one item. Current methods, though, are not able to estimate this type of importance and adjust the mining results by emphasizing the important transactions [1].
The concept of association rule mining proposes the support-confidence measurement framework and reduced association rule mining to the discovery of frequent item sets. WARM generalizes the traditional model to the case where items have weights. WARM requires for each item to be given weight to reflect their importance to the user. The weights may correspond to the profitability of different items. As more
data is gathered, which are frequently getting updated, the construction of the graph should be dynamic instead of static. Using Online Hits algorithm, the graph can be constructed dynamically and the cost can be reduced by postponing updates whenever possible. By calculating Eigen values we are enforcing the mutual reinforcement relationship between the items.

This HITS algorithm is suitable only for static content.
1) They work well in environments where no dynamic updation is possible.
2) They fail to capture the rich informations that lie within the patterns of user access or in the structure that can be defined by user group implicitly.

In this paper, we propose to replace the HITS algorithm with Online HITS algorithm which reduces the cost by postponing the updates whenever possible and makes it more suitable for Dynamic environments. HITS algorithm normally used for web pages, but it can also be used for transactional datasets from which we can calculate the hub and authorities, based on which the graph is constructed. The general HITS algorithm is too costly to run on every update. When the updates are accumulated we run online HITS once. This way cost is reduced. Another advantage of online HITS is that service of user queries, updating the Eigen vector of the given Matrix A and running of Online HITS can be made as separate activities.
The rest of the paper is arranged as follows: section II Introduction to the basic concepts of Online HITS algorithm; Section III gives an evaluation and analysis of the HITS algorithm in dynamic Environment; section IV addresses the evaluation and analysis of Online HITS algorithm in Dynamic Environments. Section V concludes the paper.

**2. Basic concept of Online HITS algorithm**

Consider the Matrix A, which is created based on the transactions and the items within each of those transactions. It will be having the values of the rules that are going to be considered for mining. The rankings say, x and y, correspond to the principal eigenvectors of the matrix $A^TA$ and $AA^T$, respectively. Even a single update to the user access will correspond to a perturbation of the A matrix. Depending on the weight function selected, it can change the behaviour of a single element or a row of elements of the matrix A. In either case the changes to the behaviour are local. These changes in behaviour may cause variations to the principal eigenvector of $A^TA$ (and $AA^T$).If we can find the relationship between the variation of x and y and the behavioural changes to the matrix A, we can check each update to see whether it will cause too much changes to x and y. If the change is within acceptable precision, we can postpone applying the update thus avoid running Online HITS for it. When the accumulated updates cause too many changes to the behaviour, we apply all the changes together and run Online HITS once. In this way the cost of running the algorithm frequently can be reduced. Such an algorithm is called as Online HITS. Another advantage of this approach is that service of user queries and updating the matrix A and running of Online HITS can be made separate. The system can update the matrix A and run Online HITS in background, and continue servicing user queries with old results that we are confident to be within certain range from the latest ones. Users can enjoy the service without any disturbances. For us, the continued accumulation of data and constructing the graph based on that is an inherent feature and the results we produce are the best guess based on the data we have so far. It is too good for the results to undergo dramatic change, which reflects the update of the latest trends and techniques about the world. Rather, we are interested in the bound of the change so that we can perform the tasks more efficiently. In addition, the conclusions are considered only to apply to unweighted graphs represented by adjacency matrices. Online HITS algorithm uses 2 main concepts, namely the computations of eigengap and perturbation. They have to be performed efficiently otherwise the cost of computing them would affect the saving of not running HITS. They will be addressed in the following subsections.

**3. Computation of Eigengap**

Eigengap is nothing but the difference between the largest and second largest Eigen values namely λ1 and λ2. The original HITS algorithm is essentially a power method to compute the principal eigenvector of S. It can be revised easily, without adding complexity, to produce λ1 and ¸ λ2 as byproducts.

Two modifications to the original HITS algorithm are introduced:

1. Find the two eigenvectors λ1 and λ2, instead of finding only the principal eigenvector. This can be done by using the block power method. Initially, start with two orthogonal vectors, multiply them all by S, then apply Gram-Schmidt to orthogonalize them. This is a single step. Iterate until they converge. In the Hits algorithm replace this step with the step that calculates principle Eigen vector.

2. HITS ensure convergence by normalizing the vector at each step to unit length. Instead, we normalize each vector by dividing them by their first non-zero element. They still converge to the two eigenvectors and the scaling factors converge to λ1 and λ2.

The above modifications introduce extra computation of one eigenvector and Gram-Schmidt orthogonalization. The Extra computation doubles the work of HITS and the order of doing orthogonalization is O(n). The total complexity is the same as HITS which is O (mn).

## 4. Evaluation and Analysis of HITS in Dynamic Environments

The usage of HITS lies in the hope that the updates may not cause too many changes (too much perturbation) to the ranking so that recomputation is avoided. In the situations where data is accumulating, running the HITS only once may not provide the correct support values since ranking of items is done only once and it is not updated frequently. The other disadvantage is that sometimes a transaction with few items may be a good hub, which is ignored by HITS. It also does not consider the difference between transactions. We have taken the customer complaints database where the updates are more frequent.

HITS Algorithm will first discover the authoritative between 2 transaction items. Given the set of n items, HITS first construct the n-by-n adjacency matrix A. The elements in row i and column j of the Matrix is 1 if there exists a relation between these 2 items, otherwise the value will be 0.

We have tested the HITS Algorithm against the Customer Complaints Database. The results of which are shown below.

Table1: Results of Complaint database using HITS Algorithm

| Complaint Type | Support Values |
|---|---|
| Complaint 1 | 0.97 |
| Complaint 2 | 0.95 |
| Complaint 3 | 0.98 |
| Complaint 4 | 0.97 |
| Complaint 5 | 0.98 |
| Complaint 6 | 0.99 |
| Complaint 7 | 1.00 |
| Complaint 8 | 0.95 |

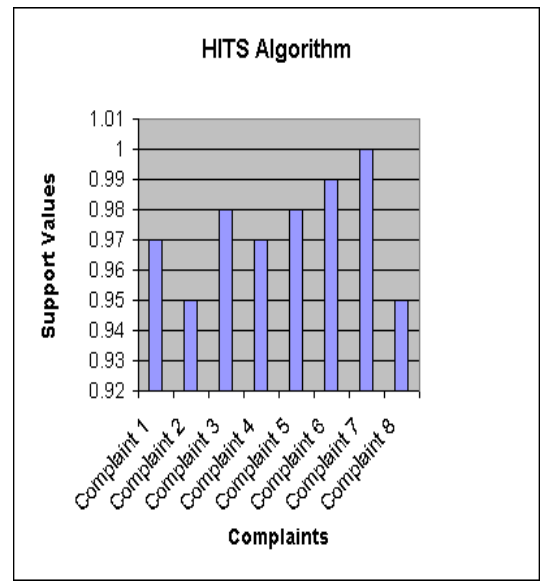

Fig 1: Graph using HITS

## 5. Evaluation and Analysis of Online HITS in Dynamic Environments

The Online HITS algorithm is nothing but a process of constructing an access based graph. Each single update of a user or every change of the item is taken to correspond to a perturbation to a matrix A. The weight function selected can perturbate a single element of the matrix A or row of elements in the matrix A. In any of these cases the Perturbation is always local. This will affect the Eigen vector of the matrix $A^TA$ and $AA^T$. If we can find the relationship between the changes of the ranking to principle eigen vector of $A^TA$ and $AA^T$ which are called as x and y values, we can check each update to see if it will cause too much changes to x and y. On verifying that the variation is within acceptable threshold level of precision, we can postpone applying the update thus avoid running Online HITS for it. Thus the Online HITS takes into account the current variations that happen dynamically, and the algorithm reflects a true dramatically changed, updated system knowledge of the current world. Online HITS constantly monitors the changes and performs operations. The results of our test are shown in the following figures.

Table2: Results of Complaint database using Online HITS Algorithm

| Complaint Type | Support Values |
|---|---|
| Complaint 1 | 0.98 |
| Complaint 2 | 0.96 |
| Complaint 3 | 0.99 |
| Complaint 4 | 0.99 |
| Complaint 5 | 0.99 |
| Complaint 6 | 0.99 |
| Complaint 7 | 1.00 |
| Complaint 8 | 0.97 |

For the above table the figure is shown as below:

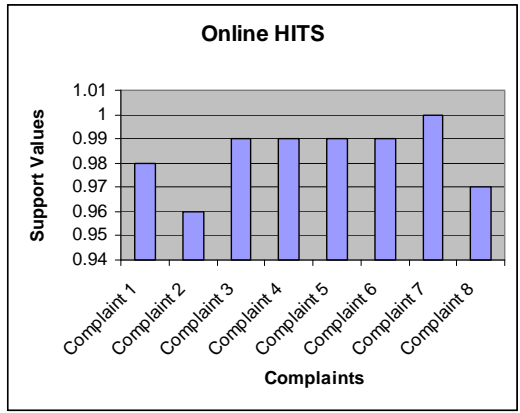

Fig2: Graph using Online HITS

Note that we are not testing how well the ranking produced by Online HITS (or HITS) fits the real ranking which is a rather qualitative and subjective measure. Instead, we are examining Online HITS's algorithmic properties and how it performs more efficient than original HITS in a dynamic system.

## 6. Conclusion

We extended the HITS hyperlink analysis algorithm to make it applicable for analyzing weighted graphs. Our generalizations are in two directions. First, we replaced the construction of static graph to the construction of dynamic graph by finding out the Eigengap. Second, we created an online eigenvector calculation method that can compute the results of mutual reinforcement voting efficiently in case of frequent updates by estimating the upper bound of perturbation and postponing applying the updates whenever possible. Ie. Till the variations are within the applicable limit, we won't run the Online HITS algorithm. Both theoretical analysis and experiments show that our generalized online algorithm is more efficient than the original HITS under the context of dynamic data.

We are going to enhance the association rule for dynamic content based on fuzzy logic and mutual reinforcement voting.

Biographies:
P.Velvadivu, B.Sc (Mathematics)-1997, M.C.A-2000, M.Phil-2002., Ph.D (Pursuing)
Places of Employment: Tata Consultancy Services : Chennai, Valtech Systems India Pvt.Ltd : Bangalore, Coimbatore Institute of Technology : Coimbatore. Member of ISTE.
Dr. K. Duraisamy, Dean, School of Science and Humanities, Kongu Engineering College, Perundurai.